\documentclass[twoside]{dis07}
\usepackage[latin1]{inputenc}
\usepackage[dvips]{graphicx,epsfig,color}
\usepackage{wrapfig,rotating}
\usepackage{amssymb,amsmath,array}

\begin{document}
\title{Nuclear $p_t$-broadening at 
 {\sc Hermes}}

\author{Yves Van Haarlem$^{(1)}$, Anton Jgoun$^{(2)}$, Pasquale Di Nezza$^{(3)}$ \\ 
on behalf of the {\sc Hermes} Collaboration
%
\vspace{.3cm}\\
%
1- Department of Subatomic and Radiation Physics, University of Gent, 9000 
Gent, Belgium \\
\vspace{.1cm}\\
2- Petersburg Nuclear Physics Institute, St. Petersburg, Gatchina, 188350 Russia \\
\vspace{.1cm}\\
3- Istituto Nazionale di Fisica Nucleare, Laboratori Nazionali di Frascati, 00044 
Frascati, Italy
}

\maketitle

\begin{abstract}
The first direct measurement of $p_t$-broadening effects
in cold nuclear matter has been studied as a function of several kinematic variables 
for different hadron types. The data have been accumulated by the {\sc Hermes} 
experiment at {\sc Desy}, in which the {\sc Hera} 27.6 GeV
lepton beam scattered off 
several nuclear gas targets.
\end{abstract}

\section{Introduction}
At {\sc Hermes} nuclear semi-inclusive deep-inelastic scattering ({\sc Sidis}) is used to study hadronization. 
In the {\sc Hermes} kinematics it is very likely that hadronization takes place inside the nucleus. In this regime the nucleus acts as a nano lab providing multiple scattering centers in the form of nucleons. Effects like the EMC effect and nuclear attenuation \cite{hermes} are already measured. An effect that is measured for the first time at {\sc Hermes} is the modification of the transverse momentum in nuclear matter or $p_t$-broadening which is presented in this work. Here, $p_t$ is the transverse momentum of the produced hadron with respect to the direction of the virtual photon. Besides the measurement of a ratio of average hadron transverse
momentum ($p_t$-ratio):
$\langle p_t^2 \rangle_A^h / \langle p_t^2 \rangle_D^h$
a new observable has been used: $\Delta \langle p_t^2 \rangle^h$, also called 
$p_t$-broadening:
\begin{equation}
\Delta \langle p_t^2 \rangle^h = \langle p_t^2 \rangle_A^h - \langle p_t^2 \rangle_D^h, 
\end{equation}
where $\langle p_t^2 \rangle_A^h$ is the average transverse momentum 
squared obtained by
a hadron of type $h$ produced on a nuclear target with atomic mass number 
$A$, and $\langle p_t^2 \rangle_D^h$ is the same but for a Deuterium target.
These measurements increase our knowledge about the space-time evolution of hadronization.

Nuclear {\sc Sidis} has the advantage that there are no initial state interactions due to the fact that leptons are point-like particles that do not contain quarks which can interact before scattering of the target. This makes the interaction easier to interprete and might help to understand the more complex heavy-ion collisions. 

$p_t$-broadening might be the most sensitive probe for
the {\it production time} as it provides a direct measurement of the production time 
$t_p$ ($\Delta \langle p_t^2 \rangle \propto t_p$) in specific models, e.g. 
\cite{kopwith}. This 
is because the hadronizing quark only contributes at time intervals $t<t_p$ to 
the $p_t$-broadening. As soon as the pre-hadron is 
formed, no further broadening occurs, because inelastic
interactions are suppressed for the pre-hadron (at $z>$ 0.5),
thus only broadening via elastic rescattering is 
still possible. Here, $z$ is the energy fraction of the virtual photon carried by the produced hadron. However, the elastic cross section is
so small that even for pions the mean free path in nuclear matter is
about $20$~fm. It is even longer for a small-size pre-hadron due to color
transparency. A disappearance of the
broadening effect is expected at large $z \to 1$ because of energy conservation.

\section{Analysis}

The data have been accumulated by the {\sc Hermes} 
experiment at {\sc Desy}, in which the {\sc Hera} 27.6 GeV positron beam scattered off 
several nuclear gas targets \cite{spec}.
Events were selected by requiring $\rm Q^2>$ 1~GeV$^2$, $\rm W^2>$ 10~GeV$^2$, and $\nu<$~23~GeV 
where $W$ is the invariant mass of the photon-nucleon system and $\nu$ is the virtual photon energy. 
Pions and Kaons are identified in the momentum range 2$<P<$15 GeV using the information from 
a ring imaging \v{C}erenkov detector.

The $p_t$-broadening effects have been studied as a function of the atomic number $\rm A$, 
$Q^2$, $\nu$, and $z$
for different hadron types produced on $\rm ^{3}He$, $\rm ^{4}He$, $\rm N$, $\rm Ne$, 
$\rm Kr$, and $\rm Xe$ targets.

The pion sample was corrected for exclusive $\rho^0$ decay pions using a
 Monte Carlo simulation. This correction was only significant in the highest 
$z$ bin where these decay pions contribute more than 50~\%. After the correction the $p_t$-ratio becomes consistent with one and the $p_t$-broadening with zero (in the highest z-bin).

The $p_t$-broadening was corrected for detector smearing, acceptance effects and QED radiative effects using a {\sc Pythia} Monte Carlo generator together with a {\sc Geant3} simulation of the {\sc Hermes} spectrometer. For $p_t$-broadening an unfolding method was used. For the $p_t$-ratio a Monte Carlo study showed that most acceptance effects cancel out except the Cahn effect which was included into the systematic uncertainty. Identified hadron samples were corrected for misidentified hadrons using an unfolding method. 

The systematic uncertainty includes contributions from the correction for 
$\rho^0$ decay pions, detector smearing and acceptance, radiative effects, and
hadron misidentification (if applicable). The dominant part in the systematic uncertainty of the $p_t$-broadening is coming from the model dependence of the acceptance correction ($\sim$ 5~\%), which is estimated using the {\sc Pythia} and the {\sc Lepto} generator, and from the Cahn effect for the $p_t$-ratio (4~\%).

\section{Results}
\begin{figure} 
  \begin{center}
    \includegraphics[width=1\linewidth]{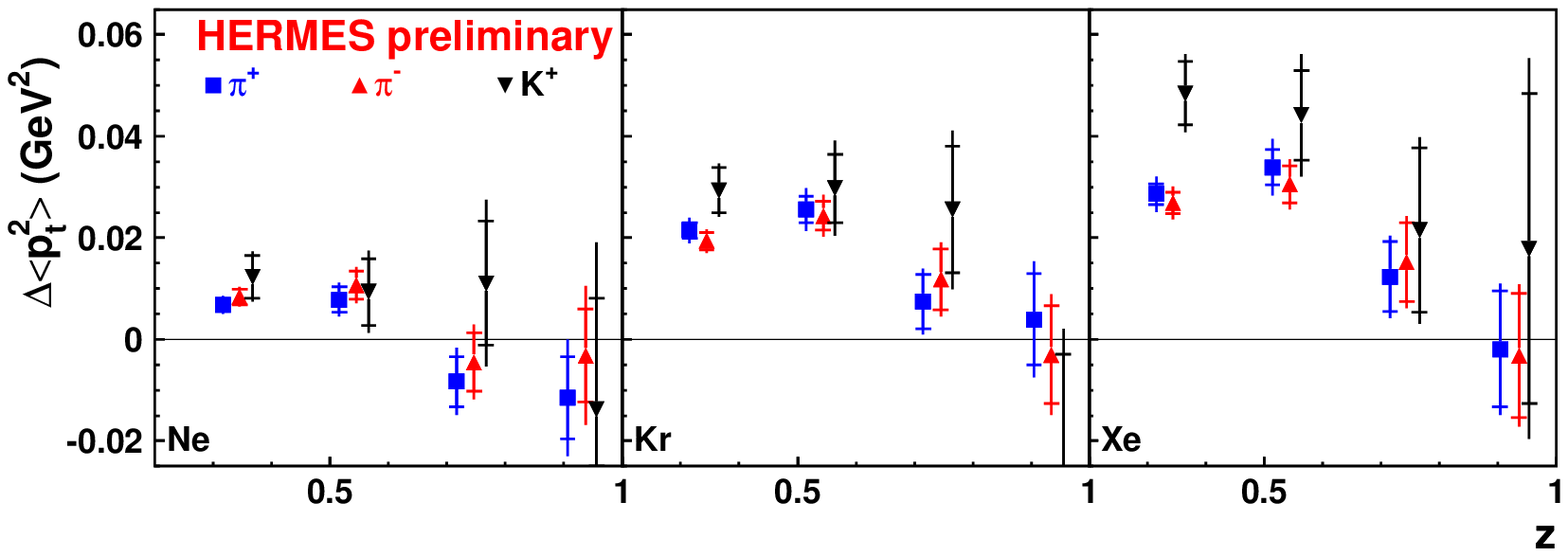}
    \includegraphics[width=1\linewidth]{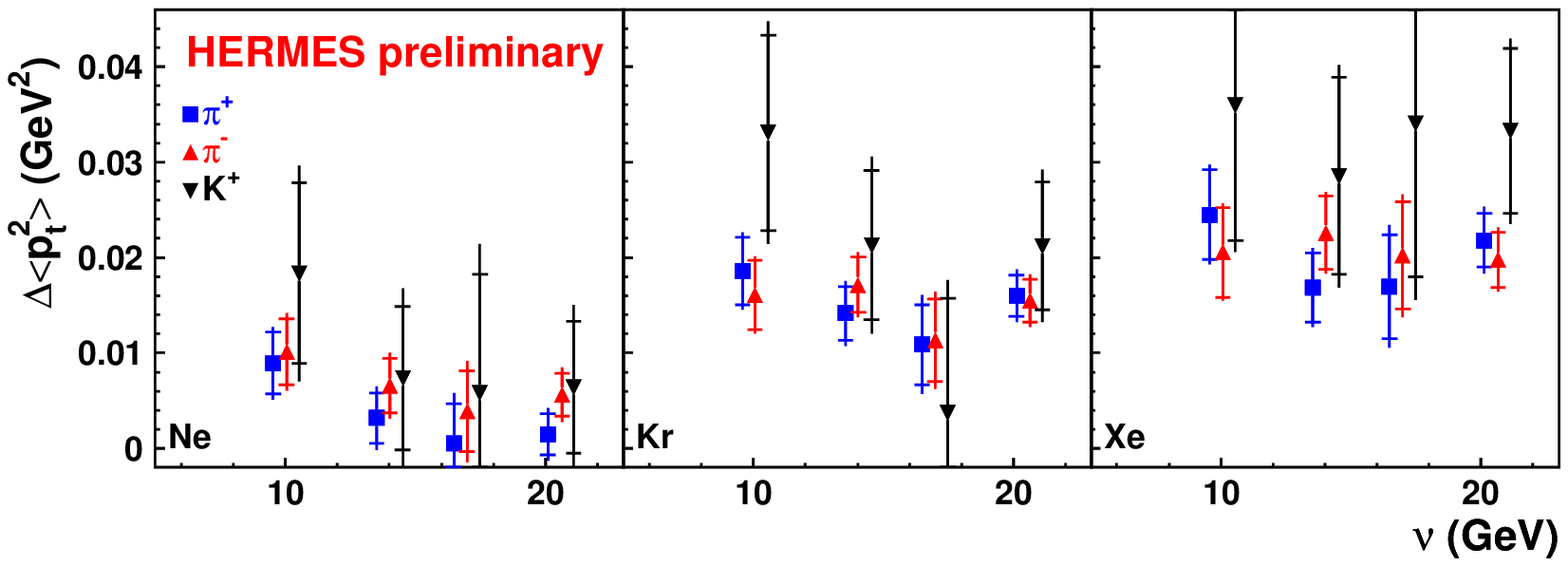}
    \includegraphics[width=1\linewidth]{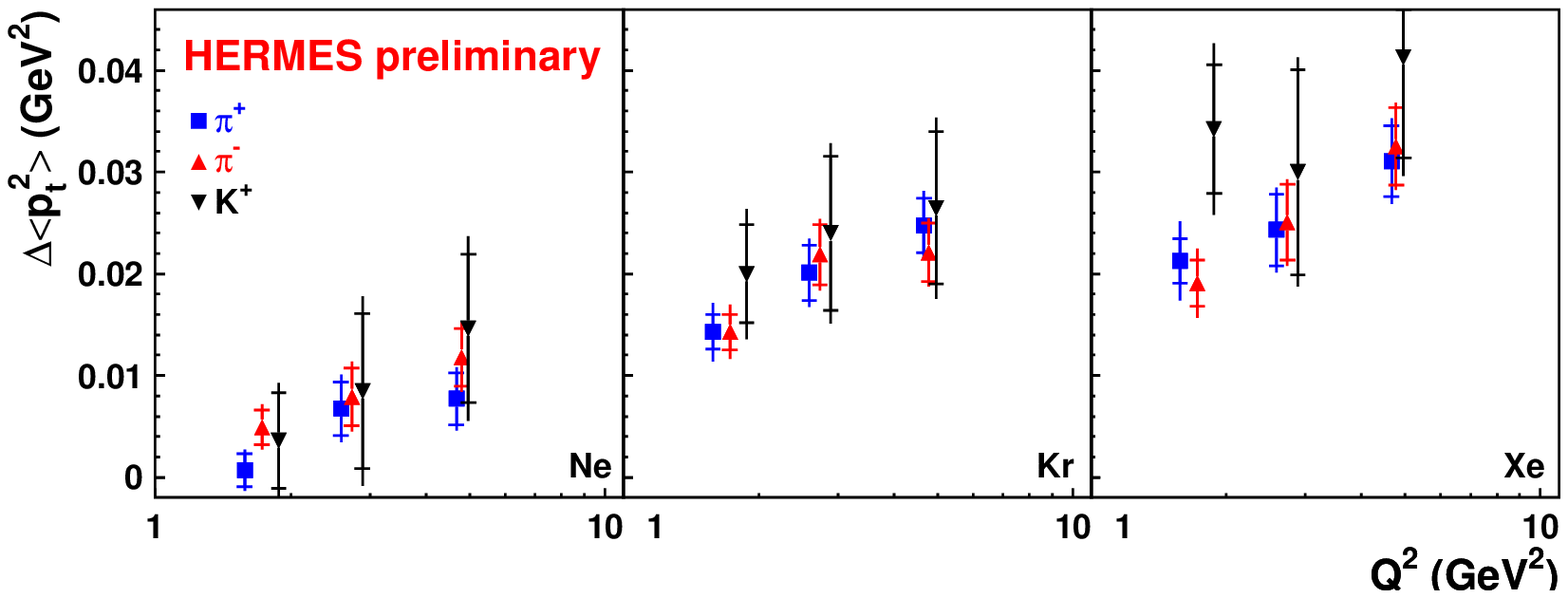}
    \caption{$p_t$-broadening for different hadron types produced from 
Ne, Kr, and Xe targets as a function of $ z$ (upper panel), $\nu$ (middle panel), and Q$^2$ (lower panel). 
The inner error bars represent the statistical error and the outer ones the quadratic sum of the
statistical and systematic uncertainties.}
   \label{ptbroad}
  \end{center}
\end{figure}

\begin{figure}
 \begin{center}
   \begin{tabular}{c c}
    \hspace{-0.8cm}\epsfig{figure=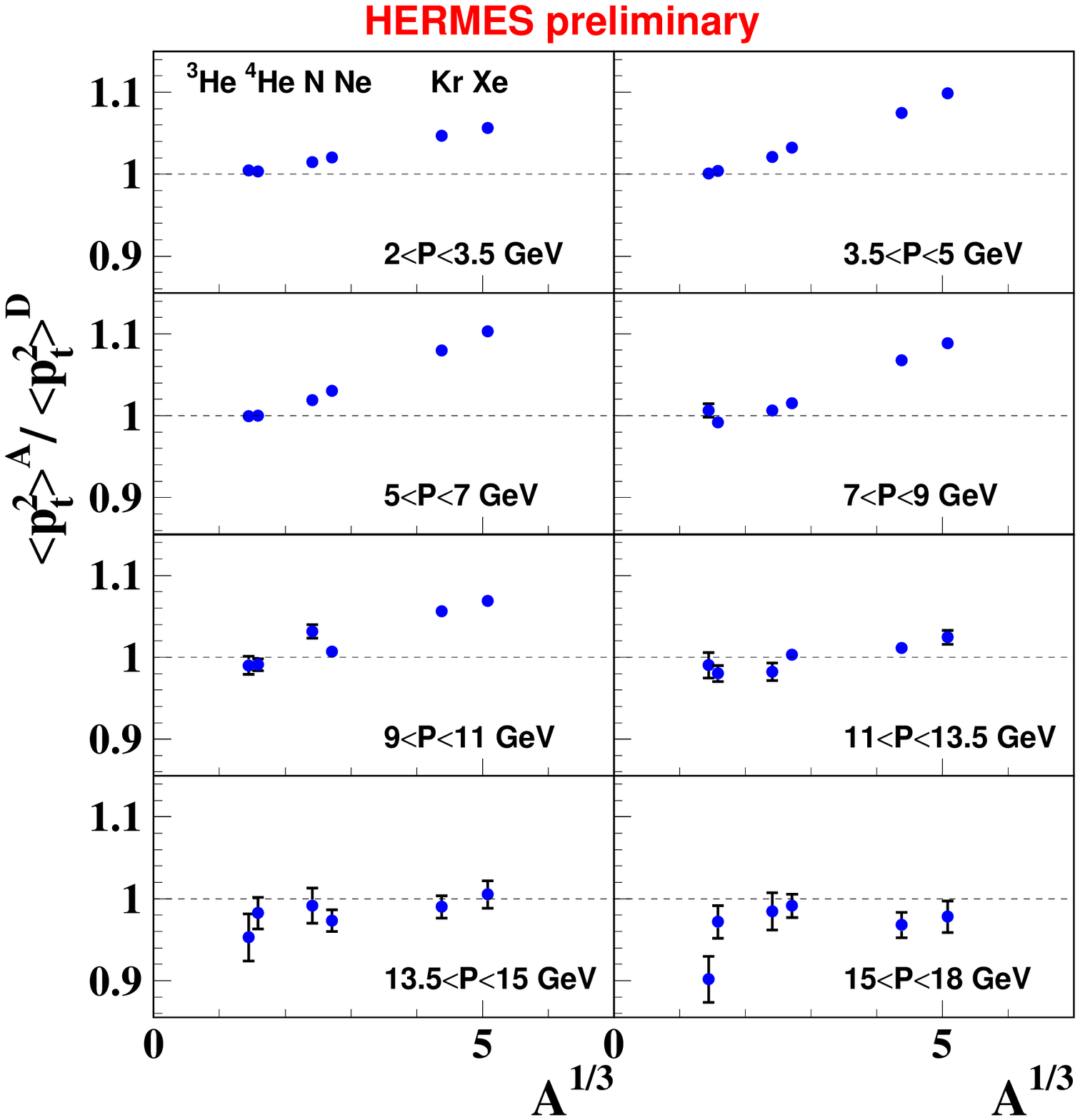,width=0.49\linewidth}
    \epsfig{figure=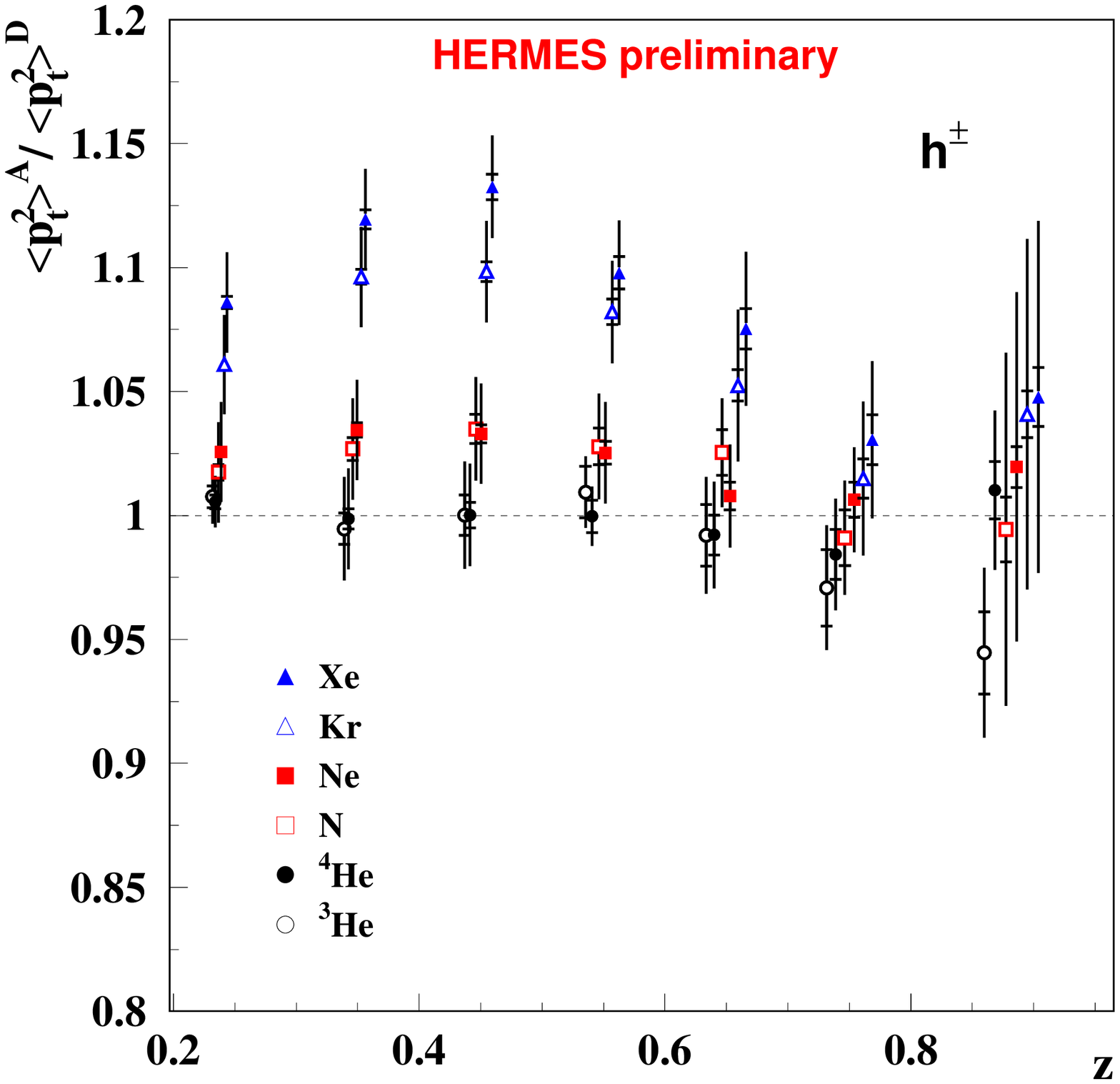,width=0.51\linewidth}
   \end{tabular}
\caption{(Left) $ p_t^2$-ratio versus A$ ^{1/3}$ for different hadron momentum 
regions for all charged hadrons. (Right) Charged hadron $ p_t$-ratio versus $ z$ for several nuclear 
targets for all charged hadrons. The inner error bars represent the statistical error and the outer ones the quadratic sum of the
statistical and systematic uncertainties.}
\label{fig:pt}
  \end{center}
\end{figure}
In figure \ref{ptbroad} (upper panel) a clear dependence of $p_t$-broadening on the atomic number A can be seen. It also shows 
that the $p_t$-broadening becomes consistent with zero as $z$ $\rightarrow$ 1.
The latter is expected by energy conservation as the fact
that a hadron with a high z value is detected means that no energy could
have been lost in any kind of interaction or reaction process. In this case the final-state hadron had to be formed immediately and $t_p~
\rightarrow~0$.
$ p_t$-broadening increases as a function of $ Q^2$, figure \ref{ptbroad} (lower panel).

Figure \ref{fig:pt} shows the $\langle p_t^2 \rangle$ ratio as a function of the atomic mass
number for different momentum ranges. For $^{3,4}He$ targets the ratio is close to one for small momenta. 
This indicates that the size of the helium nucleus is smaller than the 
hadron production time.
For heavier targets the ratio increases in the momentum range 2-7 GeV with a maximum around
7 GeV and then decreases.
The behavior of the $ p_t$-ratio for heavy targets at relative small momenta (below 7~GeV) can be caused by a production time that is smaller than the size of 
the nucleus. This could explain the increase of the $p_t$-ratio for increasing 
hadron momentum. The $p_t$-ratio decreases for high momentum,
 i.e. that the production time decreases with increasing final hadron momentum.
At very large hadron momentum there are values smaller than one. In this regime z has to be close to 1 because these hadrons have the maximum possible momentum. Such behavior 
could point out that the intrinsic momentum of the quark in a nucleon inside the nucleus is smaller than for a quark in the free nucleon.

\section{Conclusions}

The first measurement of $ p_t$-broadening effects on $^{3,4}$He, N, Ne, Kr, and Xe targets
have been presented \cite{rele}. Results were investigated for different hadron types and as a 
function of several kinematic variables. A clear signal of broadening is observed 
and it provides very important information to this physics field where a profound 
interest has been expressed by theoreticians.




%


\end{document}